\documentstyle[12pt,axodraw,epsfig,abbreviations]{article}
\def\beq{\begin{equation}}
\def\eeq{\end{equation}}
\def\bea{\begin{eqnarray}}
\def\eea{\end{eqnarray}}
\def\barr{\begin{array}}
\def\earr{\end{array}}

\begin{document}

\title{\Large\bf\boldmath
SELECTRON SEARCHES IN $e^-e^-$ SCATTERING
}
\author{{\large FRANK CUYPERS}\\\\
%\author{{FRANK CUYPERS}\\\\
{\tt cuypers@mppmu.mpg.de}\\
{\em Max-Planck-Institut f\"ur Physik, Werner-Heisenberg-Institut,}\\
{\em D-80805 M\"unchen, Germany}
}
\date{}
\maketitle

\begin{abstract}
\noindent\footnotesize
We review the pair-production and decay of selectrons
in $e^-e^-$ collisions
and show how the \sm\ backgrounds
can be virtually eliminated
with polarized beams.
The exceedingly simple analysis involved
and the large sample of background-free \susic\ events
make this \lc\ operating mode
ideal for discovering \sel s
and measuring the mass of the lightest neutralino.
\end{abstract}

\noindent
In spite of their economy of principles,
\susic\ extensions of the standard model~\cite{hkn}\
involve an opulent number of free parameters.
Although the strengths of the interactions
must be precisely the same as those of the \sm,
at this stage
the masses and mixings of the \susic\ partners
of the conventional particles
cannot be predicted from first principles.

To see clear through this jungle of parameters
it is important to perform as many independent experiments
as possible.
A particularly promising candidate
is the pair-production of \sel s in \ee\ collisions
\beq
\ee \quad\to\quad \tilde e^- \tilde e^-
\label{ss}~.
\eeq
This reaction has been studied in detail previously~\cite{ee,casc}
and we summarize here the main conclusions.

Since the strongly interacting sector
plays only a minor role
(at best)
in \ee\ collisions,
the relevant \susy\ parameters
are the mass parameters $M_1, M_2, \mu$ associated with
the $U(1)$ and $SU(2)_L$ gauginos and the higgsinos respectively,
the ratio $\tan\beta=v_2/v_1$ of the Higgs \vev s
and the selectron masses.
We work in the context of the
minimal \susic\ standard model~\cite{hkn}
and make the following assumptions:
\begin{itemize}
\item   $R$-parity is a conserved quantum number.
\item   The lightest neutralino $\tilde\chi^0_1$ is the \lsp.
\item   All selectrons have the same mass
        and are much lighter than the strongly interacting
        squarks and gluinos:
        $m_{\tilde e_L}=m_{\tilde e_R}
        \ll m_{\tilde q},m_{\tilde g}$.
\item   The mass parameters $M_1, M_2, \mu$ are real.
\item   At the GUT scale $M_1=M_2$,
        so that after renormalization to accelerator energies
        $M_1=5/3\,M_2\tan^2\theta_w$,
        where $\theta_w$ is the weak mixing angle.
\end{itemize}
The first two assumptions are essential,
because they dictate the whole \susic\ phenomenology:
all sparticles decay directly or via a cascade
into the \lsp\
which is stable and escapes detection.
These are very conservative constraints.
Indeed,
if $R$-parity were to be broken
the situation would be much simpler,
since \susy\ would show up with blatant like
lepton number violating processes or
\no\ decays in the detector.
The last three working hypotheses
are merely for simplicity
and can be relaxed without qualitatively modifying the conclusions.
Their virtue is to reduce the number of relevant independent parameters
to only four:
\beq
\tan\beta \quad M_2 \quad \mu \quad m_{\tilde\ell}
\label{param}~.
\eeq
Of these four parameters,
$\tan\beta$ is the least influential,
at least when it is larger than 2.
In contrast,
the results are very sensitive
to variations of the other three parameters.

Since the \sel\
decays only through \EW\ interactions,
its lifetime is typically long
in comparison to its mass scale,
and it is therefore
safe to use the \nwa.
Its simplest decay mode
is into an electron and the lightest \no:
\begin{equation}
        \tilde e^-\quad\to\quad e^-\No~.
\label{seldec}
\end{equation}
According to our assumptions on $R$-parity and the \lsp\ being a \no,
only the electron is visible.
If kinematically allowed,
other decays can take place like
\begin{eqnarray}
\tilde e^-
&\to&
e^-\tilde\chi^0_2
\label{seldec1}\\
&\to&
\nu_e\tilde\chi^-_1
\label{seldec2}\\
&\vdots&\nonumber
\end{eqnarray}
as well as similar decays into the heavier neutralino and chargino states.
The \susic\ particles produced this way
eventually decay into lighter (s)particles,
which themselves might undergo further decays
until only conventional particles and a number of
lightest \no s remain.
The end-product of such cascade decays
can sometimes again be an electron
accompanied by invisible particles only.
In the following
we concentrate on this very decay signature
\beq
\tilde e^-\quad\to\quad e^-+\mpT
\label{decay}~,
\eeq
whose \br\ we compute
with the two-body decay algorithm
described in Ref.~\cite{casc}.

\begin{figure}[htb]
\begin{center}
\boldmath
\unitlength1mm
\SetScale{2.837}
\begin{picture}(50,20)(-10,-30)
\ArrowLine(0,0)(15,0)
\Text(-1,0)[r]{\large$e^-$}
\ArrowLine(0,20)(15,20)
\Text(-1,20)[r]{\large$e^-$}
\Line(15,20)(15,0)
\Text(16,10)[l]{\large$\tilde\chi^0_i$}
\DashLine(15,20)(30,20){1}
\Text(31,20)[l]{\large$\tilde e^-$}
\DashLine(15,0)(30,0){1}
\Text(31,0)[l]{\large$\tilde e^-$}
%\Text(60,10)[lc]{\large$(i=1\dots4)$}
\end{picture}
\begin{picture}(80,80)(0,0)
\epsfig{file=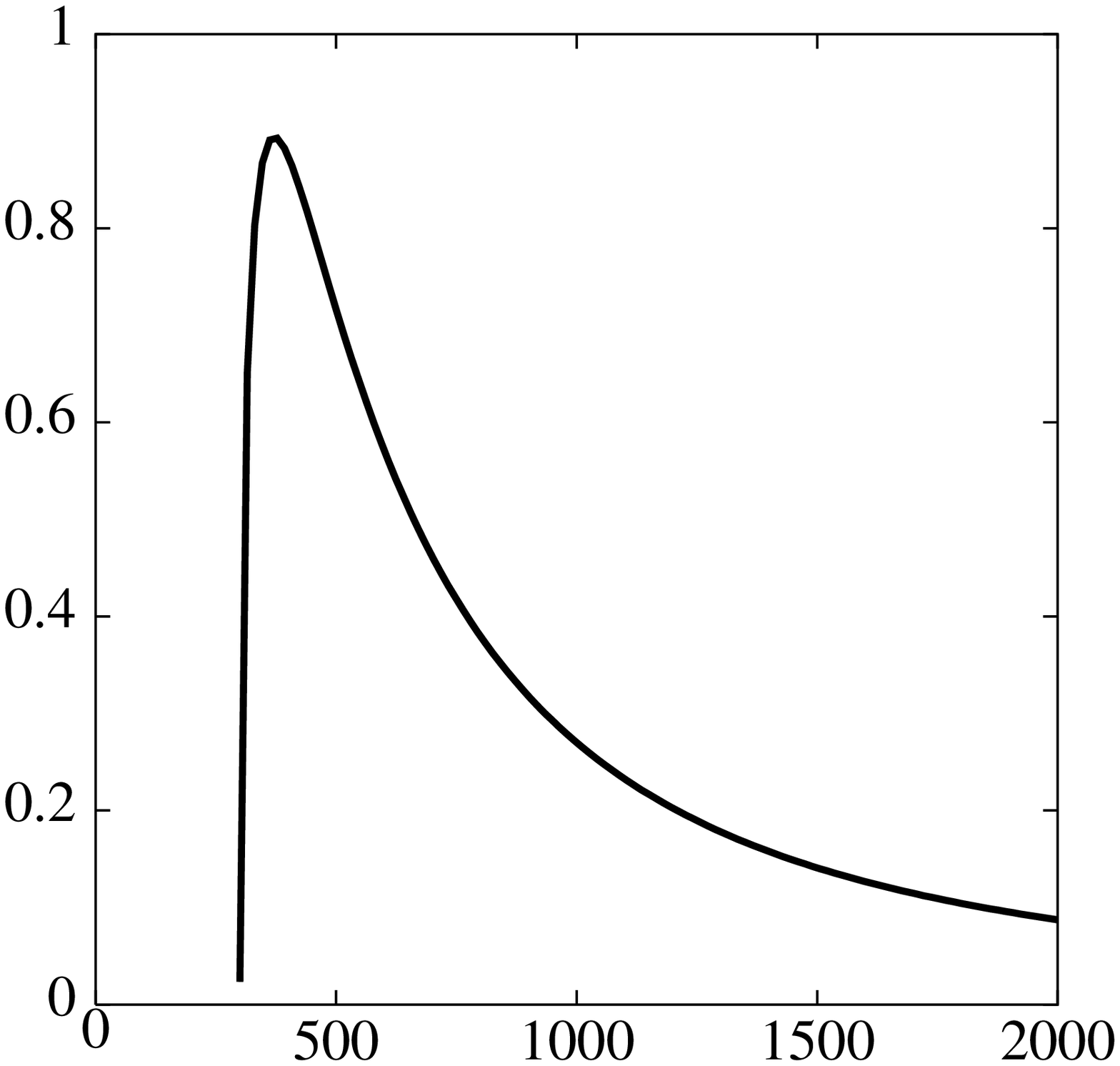,height=80mm}
\Text(-80,75)[tr]{\large$\sigma$ [pb]}
\Text(0,0)[tr]{\large$\sqrt{s_{ee}}$ [GeV]}
\Text(-40,70)[tl]{\fbox{\normalsize $e^-e^- \to \tilde e^-\tilde e^-$}}
\Text(-40,60)[tl]{\normalsize $\tan\beta = 10$}
\Text(-40,55)[tl]{\normalsize $\mu = -300$ GeV}
\Text(-40,50)[tl]{\normalsize $M_2 = 200$ GeV}
\Text(-40,45)[tl]{\normalsize $m_{\tilde\ell} = 150$ GeV}
\end{picture}
\end{center}
\caption[f5]{\footnotesize
  Lowest order Feynman diagram
  describing \sel\ production in $e^-e^-$ collisions
  and typical energy dependence of the corresponding \xs.}
\label{f5}
\end{figure}

Selectron pair-production takes place in \ee\ collisions
via the exchange of \no s,
as depicted in Fig.~\ref{f5}.
Note that all four \no s play an important role
in this reaction~\cite{casc}.
The dependence on the \susy\ parameters is thus rather complex
because it enters at three different levels:
\begin{enumerate}
\item through the masses of the four different \no s;
\item through their mixings among each other which affects
  their couplings to electrons and \sel s;
\item through the mass and \br s of the \sel.
\end{enumerate}
The energy dependence of the \xs\
is also shown in Fig.~\ref{f5},
for unpolarized beams.
For this
we imposed the following rapidity, energy and acoplanarity cuts
on the observed leptons:
\begin{equation}
|\eta_e| < 3
\quad , \quad
E_e > 5 \GeV
\quad , \quad
||\phi(e^-_1)-\phi(e^-_2)|-180^\circ| > 2^\circ
\label{cut}\ ,
\end{equation}
where $\phi$ is the azimuthal angle of the decay electrons with
respect to the beam axis.
The yield is sharply peaked just above threshold
so that an energy scan can provide precise information
about the mass of the \sel.

Concentrating on the \sel\ decay Eq.~(\ref{decay})
leads to the following observable signal:
\beq
\ee \quad\to\quad \tilde e^-\tilde e^- \quad\to\quad \ee+\mpT
\label{signal}~.
\eeq
The most important \sm\ backgrounds
originate from $W^-$ and $Z^0$ Brems\-strah\-lung
{\arraycolsep0cm
\renewcommand{\arraystretch}{0}
\bea
e^-e^-
& \quad\to\quad &
\begin{array}[t]{ll}
  e^-\nu_e&W^- \qquad \\
  &\hra e^-\bar\nu_e
\end{array}
\label{w}\\\nonumber\\
e^-e^-
& \quad\to\quad &
\begin{array}[t]{ll}
  e^-e^-&Z^0 \qquad \\
  &\strut\hra\nu\bar\nu
\end{array}
\label{z}
\eea
}

\noindent
After imposing the acceptance cuts (\ref{cut})
and including the relevant branching ratios,
the \xs s for $\sqrt{s_{ee}}=500$ GeV
are 150 fb for $W^-$
and 40 fb for $Z^0$ Bremsstrahlung~\cite{ee}.
The potential background from M\o ller scattering
is entirely eliminated by the acoplanarity cut.
The \susic\ signal,
on the other hand,
is not significantly reduced by these mild cuts,
which roughly simulate a typical detector acceptance.
As a result,
over all the kinematically accessible \susy\ parameter space
({\em i.e.} if the collider energy is sufficient to pair-produce the \sel s)
the signal-to-background ratio is never significantly less than one.
Simply counting the rates is thus largely sufficient to discover the \sel s.

\begin{figure}[htb]
\begin{center}
\boldmath
\unitlength1mm
\SetScale{2.837}
\begin{picture}(110,85)(-10,0)
\includegraphics{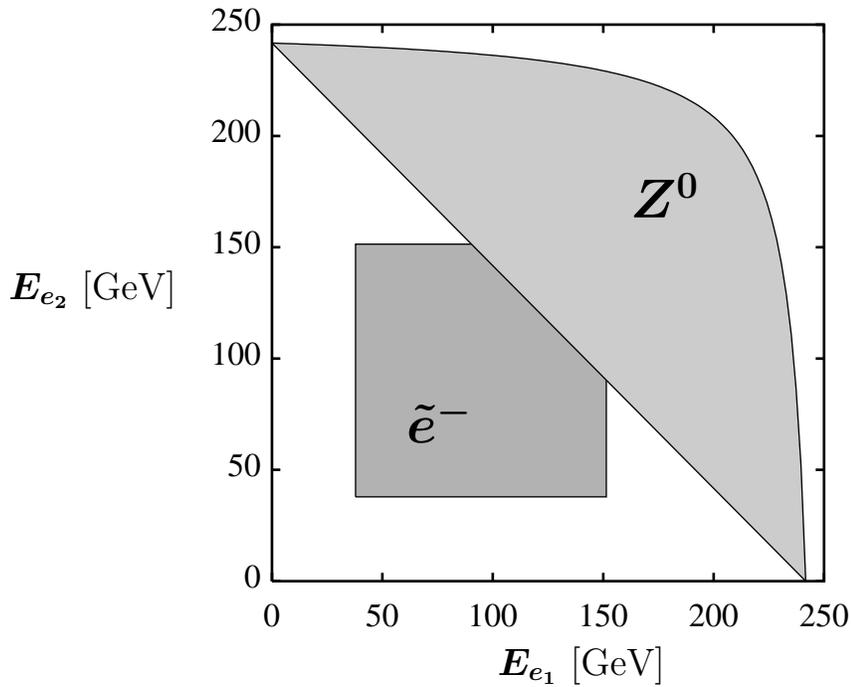}
\Text(5,50)[tr]{\large $E_{e_2}$ [GeV]}
\Text(70,0)[tr]{\large $E_{e_1}$ [GeV]}
\Text(40,30)[c]{\LARGE {$\tilde e^-$}}
\Text(70,60)[c]{\LARGE {$Z^0$}}
\end{picture}
\end{center}
\caption{\footnotesize
  Dalitz plot of the allowed energy ranges
  of the final state electrons in the processes
  $e^-e^-\to e^-e^-Z^0$
  (diagonal line and hyperbola)
  and
  $e^-e^-\to\tilde e^-\tilde e^-\to e^-e^-\tilde\chi^0_1\tilde\chi^0_1$.
  (square).
  For the latter reaction
  we have assumed $m_{\tilde e}=$ 200 GeV
  and $m_{\tilde \chi^0_1}=$ 100 GeV.
}
\label{dalitz}
\end{figure}

\begin{figure}[htb]
\begin{center}
\boldmath
\unitlength1mm
\SetScale{2.837}
\begin{picture}(50,0)(0,-60)
\Text(0,0)[tl]{\normalsize \fbox{$e_R^-e_R^- \to e^-e^-+\mpT$}}
\Text(0,-15)[tl]{\normalsize $\sqrt{s_{ee}} = 500$ GeV}
\Text(0,-20)[tl]{\normalsize $\tan\beta = 10$}
\Text(0,-25)[tl]{\normalsize $m_{\tilde\ell} = 200$ GeV}
\end{picture}
\begin{picture}(80,80)(0,0)
\epsfig{file=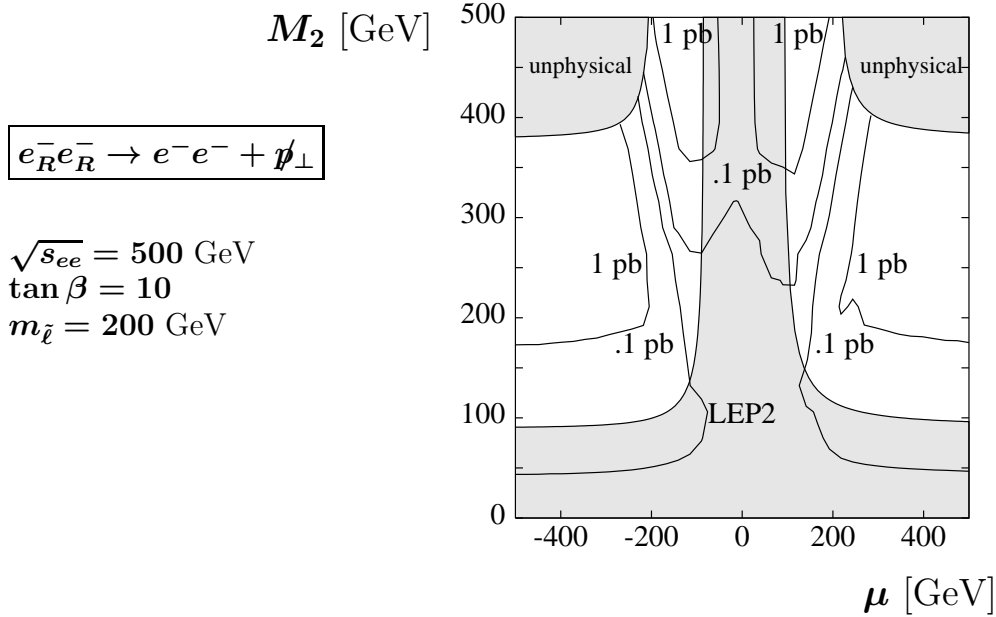,height=80mm}
\Text(-75,75)[tr]{\large $M_2$ [GeV]}
\Text(0,0)[tr]{\large $\mu$ [GeV]}
\end{picture}
\end{center}
\caption[f8]{\footnotesize
  Contours in the \susy\ parameter space
  of constant \xs s for the selectron signal.
  The cuts (\ref{cut},\ref{e1e2cut}) are included.
  The regions where $m_{\tilde e} < m_{\tilde\chi^0_1}$
  are excluded by assumption
  and are labeled ``unphysical''.
%  The contours which would be obtained if cascade decays are ignored
%  are also shown with dotted lines.
}
\label{f8}
\end{figure}

The signal to background ratio
can be strongly enhanced with right-handed electron beams,
for which the $W^-$ Bremsstrahlung background (\ref{w}) disappears.
It is then worthwhile to also eliminate the background
from $Z^0$ Bremsstrahlung,
in order to select a clean sample of \susic\ events
with no,
or negligibly little,
background from \sm\ processes.
As can be gathered from Fig.~\ref{dalitz},
where we plotted the kinematically admissible energies
of the emerging electrons,
the $Z^0$  events can be filtered out
by rejecting all \ee\ events
with a total deposited energy exceeding about half the \cm\ energy
\begin{equation}
E_{e_1}+E_{e_2} ~<~ {s-m_Z^2\over2\sqrt{s}}% ~\approx~ 242 \GeV
\label{e1e2cut}\ .
\end{equation}
If this cut is imposed,
none of the $Z^0$ contributes
and at worst 55\%\ of the electron pairs
which originate from \sel\ production are lost.
The next order irreducible background then
originates from double $W^-$ Bremsstrahlung %(\ref{ww})
{\arraycolsep0cm
\renewcommand{\arraystretch}{0}
\begin{equation}
        e^-e^- \quad\to\quad
        \begin{array}[t]{ll}
                W^-\nu_e&W^-\nu_e \qquad ,\\
                        &\hra e^-\bar\nu_e\\
                \multicolumn{2}{l}{\hra e^-\bar\nu_e}
        \end{array}
\label{ww}
\end{equation}
}

\noindent
and amounts to about .1 fb at 500 GeV~\cite{ckr}.
We have plotted in Fig.~\ref{f8}
the contours in the $(\mu,M_2)$ half-plane
along which the observable \xs\ for the $e^-e^-+\mpT$ signal
from the production of 200 GeV \sel s
is 1 and 0.1 pb.

This nearly total absence of backgrounds
is a unique opportunity
for performing studies
which would be arduous or impossible
in any other environment,
like \ep\ annihilations.
In particular,
one can think of
the measurement of the \no\ mass~\cite{ee} and
the study of cascade decays~\cite{casc},
to which we turn now.

The mass of the lightest \no\ can be determined kinematically
from the endpoints $E_{\rm min,max}$
of the electron energy distribution:
\bea
m_{\tilde\chi^0_1}^2
&=&
\sqrt{s_{ee}}
{E_{\rm max}E_{\rm min} \over E_{\rm max}+E_{\rm min}}
\left( {\sqrt{s_{ee}} \over E_{\rm max}+E_{\rm min}} -2 \right)
\label{e100}\ .
\eea
This is a totally model-independent measurement of the mass of the \lsp,
which no other experiment can perform as precisely.
Of course,
there is always some smearing due to initial state Bremsstrahlung
and beamstrahlung,
and the incidence of these effects should be further investigated.

Softer electrons emerging at the end
of a longer cascade such as the ones initiated by the
decays (\ref{seldec1},\ref{seldec2})
will not be very much affected by the cuts (\ref{cut},\ref{e1e2cut}).
This makes the \ee\ \lc\ mode
an ideal and unique tool
for observing and studying \susic\ cascades.
Neither hadronic nor \ep, \eg\ or \pp\ collisions
can perform well in this field,
because they all require high \trm\ cuts
in order to enhance the signal to background ratio.

To conclude,
we have shown that the \ee\ operating mode of a \lc\
is ideal for discovering and studying \sel s.
This is mainly due to the low \sm\ activity of \ee\ collisions,
which provides a low background environment.
Kinematically accessible \sel s
would be revealed by a simple counting experiment.
Moreover,
an extremely pure sample of right-\sel s can be obtained
with polarized beams.
This in turn allows
the precise measurement of the \no\ mass and
the opportunity to observe and analyze cascade decays of the \sel.

\end{document}